\documentclass{snmp2003}

\renewcommand{\thefootnote}{\fnsymbol{footnote}}
\begin{document}
\ShortArticleName{Family Replicated Gauge Group Models} 
\ArticleName{Family Replicated Gauge Group Models\footnote[1]{Talk 
given by L.~V.~Laperashvili at the Fifth International 
Conference ``Symmetry in Nonlinear Mathematical Physics'', 
Kyiv, Ukraine, June 23-29, 2003.}}
\Author{C.~D.~Froggatt~$^\dag$, L.~V.~Laperashvili~$^\ddag$,
H.~B.~Nielsen~$^\S$, Y.~Takanishi~$^\P$}
\AuthorNameForHeading{C.~D.~Froggatt, L.~V.~Laperashvili, 
H.~B.~Nielsen, Y.~Takanishi}


\Address{$^\dag$~Department of Physics and Astronomy, Glasgow University, 
Glasgow G12 8QQ, Scotland}
\EmailD{c.froggatt@physics.gla.ac.uk}
\Address{$^\ddag$~ITEP, 
B.~Cheremushkinskaya 25, 117218 Moscow, Russia}
\EmailD{laper@heron.itep.ru}
\Address{$^\S$~The Niels Bohr Institute, Blegdamsvej 17, 
2100 Copenhagen {\O}, Denmark}
\EmailD{hbech@alf.nbi.dk}
\Address{$^\P$~The Abdus Salam ICTP, Strada Costiera 11, 
34100 Trieste, Italy}
\EmailD{yasutaka@ictp.trieste.it}

\Abstract{{} Family Replicated Gauge Group models of the type
  $SU(n)^N\times SU(m)^N$, $(SMG)^3$ and $(SMG\times U(1)_{B-L})^3$
  are reviewed, where $SMG=SU(3)_{c}\times SU(2)_{L}\times U(1)_{Y}$
  is the gauge symmetry group of the Standard Model, $B$ is the baryon
  and $L$ is the lepton numbers, respectively. It was shown that
  Family Replicated Gauge Group model of the latter type fits the
  Standard Model fermion masses and mixing angles and describes all
  neutrino experiment data order magnitudewise using only 5 free
  parameters -- five vacuum expectation values of the Higgs fields
  which break the Family Replicated Gauge Group symmetry to the
  Standard Model. The possibility of $[SU(5)]^3$ or $[SO(10)]^3$
  unification at the GUT-scale $\sim 10^{18}$ GeV also is briefly
  considered.}

\renewcommand{\thefootnote}{\arabic{footnote}}
\setcounter{footnote}{0}
\section{Introduction}

Trying to gain insight into Nature and considering the physical
processes at very small distances, physicists have made attempts to
explain the well-known laws of low-energy physics as a consequence of
the more fundamental laws of Nature. The contemporary low-energy
physics of the electroweak and strong interactions is described by the
Standard Model (SM) which unifies the Glashow-Salam-Weinberg
electroweak theory with QCD -- the theory of strong interactions.

The gauge symmetry group in the SM is :
\begin{equation}
  SMG = SU(3)_{c}\times SU(2)_{L}\times U(1)_{Y}\,,
  \label{1}
\end{equation}
which describes elementary particle physics up to the scale
$\approx 100$ GeV.

Recently it was shown that the Family Replicated Gauge Groups (FRGG)
of the type $SU(n)^N\times SU(m)^N$ provide new directions for
research in high energy physics and quantum field theory. In the
Deconstruction of space-time models~\cite{Laperashvili:1}, the authors
tried to construct renormalizable asymptotically free 4-dimensional
gauge theories which dynamically generate a fifth dimension (it is
possible to obtain more dimensions in this way).  Such theories
naturally lead to electroweak symmetry breaking, relying neither on
supersymmetry nor on strong dynamics at the TeV scale. The new TeV
physics is perturbative and radiative corrections to the Higgs mass
are finite. Thus, we see that the family replicated gauge groups
provide a new way to stabilize the Higgs mass in the Standard Model.

But there exists quite different way to employ the FRGG.

\section{Family Replicated Gauge Group as an extension of the Standard Model}

The extension of the Standard Model with the Family Replicated Gauge
Group :
\begin{equation}
G = (SMG)^{N_{fam}} = [SU(3)_c]^{N_{fam}}\times
[SU(2)_L]^{N_{fam}} \times [U(1)_Y]^{N_{fam}}       
\label{2}
\end{equation}
was first suggested in the paper~\cite{Laperashvili:2} and developed
in the book~\cite{Laperashvili:3} (see also the
review~\cite{Laperashvili:4}). Here $N_{fam}$ designates the number of
quark and lepton families. If $N_{fam}=3$ (as our theory predicts and
experiment confirms), then the fundamental gauge group $G$ is:
\begin{equation}
G = (SMG)^3 = SMG_{1st\;fam.}\times SMG_{2nd\;fam.}\times SMG_{3rd\;fam.},
\label{3}
\end{equation}
or
\begin{equation}
G = (SMG)^3 = {[SU(3)_c]}^3\times {[SU(2)_L]}^3\times {[U(1)_Y]}^3.
\label{4}
\end{equation}
The generalized fundamental group:
\begin{equation}
  G_f = (SMG)^3\times U(1)_f 
\label{5}
\end{equation}
was suggested by fitting the SM charged fermion masses and mixing
angles in paper~\cite{Laperashvili:5}. A new generalization of our
FRGG-model was suggested in papers~\cite{Laperashvili:6}, where:
\begin{eqnarray}
G_{\mbox{ext}} \!&=&\! (SMG\times U(1)_{B-L})^3 \nonumber \\
\!&\equiv&\! [SU(3)_c]^3\times [SU(2)_L]^3\times [U(1)_Y]^3
\times [U(1)_{(B-L)}]^3
\label{6}
\end{eqnarray}
is the fundamental gauge group, which takes right-handed neutrinos
and the see-saw mechanism into account. This extended model can
describe all modern neutrino experiments, giving a reasonable fit to
all the quark-lepton masses and mixing angles.

The gauge group $G=G_{ext}$ contains: $3\times8=24$ gluons, 
$3\times3=9$ $W$-bosons, and $3\times1+3\times1=6$ Abelian gauge bosons.

The gauge group $G_{\mbox{ext}}=(SMG\times U(1)_{B-L})^3$ undergoes
spontaneous breakdown (at some orders of magnitude below the Planck
scale) to the Standard Model Group SMG which is the diagonal subgroup
of the non-Abelian sector of the group $G_{\mbox{ext}}$. As was shown
in Ref.~\cite{Laperashvili:7}, 6 different Higgs fields: $\omega$,
$\rho$, $W$, $T$, $\phi_{WS}$, $\phi_{B-L}$ break our FRGG-model to
the SM. The field $\phi_{WS}$ corresponds to the Weinberg-Salam Higgs
field of Electroweak theory. Its vacuum expectation value (VEV) is
fixed by the Fermi constant: $\left\langle\phi_{WS}\right\rangle=246$
GeV, so that we have only 5 free parameters -- five VEVs:
$\left\langle\omega\right\rangle$, $\left\langle\rho\right\rangle$,
$\left\langle W \right\rangle$, $\left\langle T \right\rangle$,
$\left\langle \phi_{B-L} \right\rangle$ to fit the experiment in the
framework of the SM. These five adjustable parameters were used with
the aim of finding the best fit to experimental data for all fermion
masses and mixing angles in the SM, and also to explain the neutrino
oscillation experiments.

Experimental results on solar neutrino and atmospheric neutrino
oscillations from Sudbury Neutrino Observatory (SNO Collaboration) and
the Super-Kamiokande Collaboration have been used to extract the
following parameters:
\begin{equation}
\Delta m^2_{\rm solar} = m_2^2 - m_1^2,\quad\Delta m^2_{\rm atm} = m_3^2 -
m_2^2,\quad\tan^2 \theta_{\rm solar} = \tan^2 \theta_{12},\quad\tan^2
\theta_{\rm atm} = \tan^2 \theta_{23}\, 
\end{equation} 
where $m_1,\,m_2,\,m_3$ are the hierarchical left-handed neutrino
effective masses for the three families. We also use the CHOOZ reactor
results. It is assumed that the fundamental Yukawa couplings in our
model are of order unity and so we make order of magnitude
predictions. The typical fit is shown in Table I. As we can see, the 5
parameter order of magnitude fit is encouraging.
%
\begin{center}
\begin{table}[!th]
\caption{Best fit to conventional experimental data.
All masses are running masses at $1~\mbox{\rm GeV}$ 
except the top quark mass which is the pole mass.}
\begin{displaymath}
\begin{array}{|c|c|c|}
\hline 
 & {\rm Fitted} & {\rm Experimental} \\ \hline
m_u & 4.4~\mbox{\rm MeV} & 4~\mbox{\rm MeV} \\ \hline
m_d & 4.3~\mbox{\rm MeV} & 9~\mbox{\rm MeV} \\ \hline
m_e & 1.6~\mbox{\rm MeV} & 0.5~\mbox{\rm MeV} \\ \hline
m_c & 0.64~\mbox{\rm GeV} & 1.4~\mbox{\rm GeV} \\ \hline
m_s & 295~\mbox{\rm MeV} & 200~\mbox{\rm MeV} \\ \hline
m_{\mu} & 111~\mbox{\rm MeV} & 105~\mbox{\rm MeV} \\ \hline
M_t & 202~\mbox{\rm GeV} & 180~\mbox{\rm GeV} \\ \hline
m_b & 5.7~\mbox{\rm GeV} & 6.3~\mbox{\rm GeV} \\ \hline
m_{\tau} & 1.46~\mbox{\rm GeV} & 1.78~\mbox{\rm GeV} \\ \hline
V_{us} & 0.11 & 0.22 \\ \hline
V_{cb} & 0.026 & 0.041 \\ \hline
V_{ub} & 0.0027 & 0.0035 \\ \hline
\Delta m^2_{\odot} & 9.0 \times 10^{-5}~\mbox{\rm eV}^2 &  
5.0 \times 10^{-5}~\mbox{\rm eV}^2 \\ \hline
\Delta m^2_{\rm atm} & 1.7 \times 10^{-3}~\mbox{\rm eV}^2 &  2.5 
\times 10^{-3}~\mbox{\rm eV}^2\\ \hline
\tan^2\theta_{\odot} &0.26 & 0.34\\ \hline
\tan^2\theta_{\rm atm}& 0.65 & 1.0\\ \hline
\tan^2\theta_{\rm chooz}  & 2.9 \times 10^{-2} & <2.6 \times 10^{-2}\\
\hline 
\end{array}
\end{displaymath}
\end{table} %
\end{center}
%
There are also 3 see-saw heavy neutrinos in this model (one
right--handed neutrino in each family) with masses:
$M_1,\,M_2,\,M_3.\,$ The model predicts the following neutrino
masses: 
\begin{equation} m_1\approx 1.4\times 10^{-3} \; {\mbox{eV}},\quad
m_2\approx 9.6\times 10^{-3} \; {\mbox{eV}},\quad m_3\approx
4.2\times 10^{-2} \; {\mbox{eV}}
\end{equation} 
-- for left-handed neutrinos, and 
\begin{equation} 
M_1\approx 1.0\times 10^6 \; {\mbox{GeV}},\quad M_2\approx
6.1\times 10^9 \; {\mbox{GeV}}, \quad M_3\approx 7.8\times 10^9 \;
{\mbox{GeV}} 
\end{equation} 
-- for right-handed (heavy) neutrinos.

Finally, we conclude that our theory with the FRGG-symmetry is
very successful in describing experiment. The best fit gave the
following values for the VEVs:
\begin{equation}
  \left\langle W\right\rangle\approx 0.157,\quad 
  \left\langle T \right\rangle\approx 0.077,\quad
   \left\langle\omega\right\rangle\approx 0.244,\quad 
\left\langle\rho\right\rangle\approx 0.265
\end{equation} 
in the ``fundamental units'', $M_{Pl}=1$, and
\begin{equation}
  \left\langle\phi_{B-L}\right\rangle
\approx 5.25\times 10^{15} \,\,{\mbox{GeV}} \,
\end{equation} which gives the see-saw scale: the scale of breakdown of the
$U(1)_{B-L}$ groups ($\sim 5\times 10^{15}\,$ GeV).

\section{The Problem of Monopoles in the Standard and Family 
Replicated Models}

The aim of the present Section is to show, following 
Ref.~\cite{Laperashvili:8}, that
monopoles cannot be seen in the Standard Model and in its usual
extensions in the literature up to the Planck scale:
$M_{Pl}=1.22\times 10^{19}$ GeV, because they have a huge
magnetic charge and are completely confined or screened.
Supersymmetry does not help to see monopoles.

In theories with the FRGG-symmetry the charge of monopoles is
essentially diminished. Then monopoles can appear near the Planck
scale and change the evolution of the fine structure constants
$\alpha_i(t)$ (here $i=1,2,3$ corresponds to $U(1)$, $SU(2)$ and $SU(3)$),
$t=\log(\mu^2/\mu_R^2)$, where $\mu$ is the 
energy variable and $\mu_R$ is the renormalisation point.

Let us consider the ``electric'' and ``magnetic'' fine structure
constants:
\begin{equation}
    \alpha = \frac{g^2}{4\pi}\quad{\mbox{and}}\quad
    \tilde \alpha = \frac{\tilde g^2}{4\pi},                   
\label{10m}
\end{equation}
where $g$ is the coupling constant, and $\tilde g$ is the dual
coupling constant (in QED: $g=e\,({\mbox{electric charge}})$,
and $\tilde g = m \, ({\mbox{magnetic charge}})$).

The Renormalization Group Equation (RGE) for monopoles is: 
\begin{equation}
\frac {d(\log \tilde \alpha(t))}{dt} = \beta(\tilde \alpha).
\label{16m}
\end{equation} 
With the scalar monopole beta-function we have: 
\begin{equation}
   \beta(\tilde \alpha) = \frac{\tilde \alpha}{12\pi} +
{(\frac{\tilde \alpha}{4\pi })}^2 + \cdots 
= \frac{\tilde\alpha}{12\pi }( 1 +
 3\frac{\tilde \alpha}{4\pi } + \cdots)\,.                
\label{1m}
\end{equation} 
The last equation shows that the theory of monopoles cannot be
considered perturbatively at least for
\begin{equation}
\tilde \alpha > \frac{4\pi}{3}\approx 4.
\label{2m}
\end{equation} %
And this limit is smaller for non-Abelian monopoles.

Let us consider now the evolution of the SM running fine structure
constants. The usual definition of the SM coupling constants is
given in {\it the Modified minimal subtraction scheme} ($\it\overline{\rm MS}$): 
\begin{equation}
  \alpha_1 = \frac{5}{3}\alpha_Y,\quad
  \alpha_Y = \frac{\alpha}{\cos^2\theta_{\overline{\rm MS}}}\,,\quad
  \alpha_2 = \frac{\alpha}{\sin^2\theta_{\overline{\rm MS}}}\,,\quad
  \alpha_3 \equiv \alpha_s = \frac {g^2_s}{4\pi} \,,
\end{equation} 
where $\alpha$ and $\alpha_s$ are the electromagnetic and
$SU(3)$ fine structure constants respectively, $Y$ is the weak
hypercharge, and $\theta_{\overline{\rm MS}}$ is the Weinberg weak angle in
$\overline{\rm MS}$ scheme. Using RGEs with experimentally established
parameters, it is possible to extrapolate the experimental values
of the three inverse running constants $\alpha_i^{-1}(\mu)$ from
the Electroweak scale to the Planck scale (see Fig.~1).

\begin{figure}[t]
\centerline{\epsfxsize=\textwidth \epsfbox{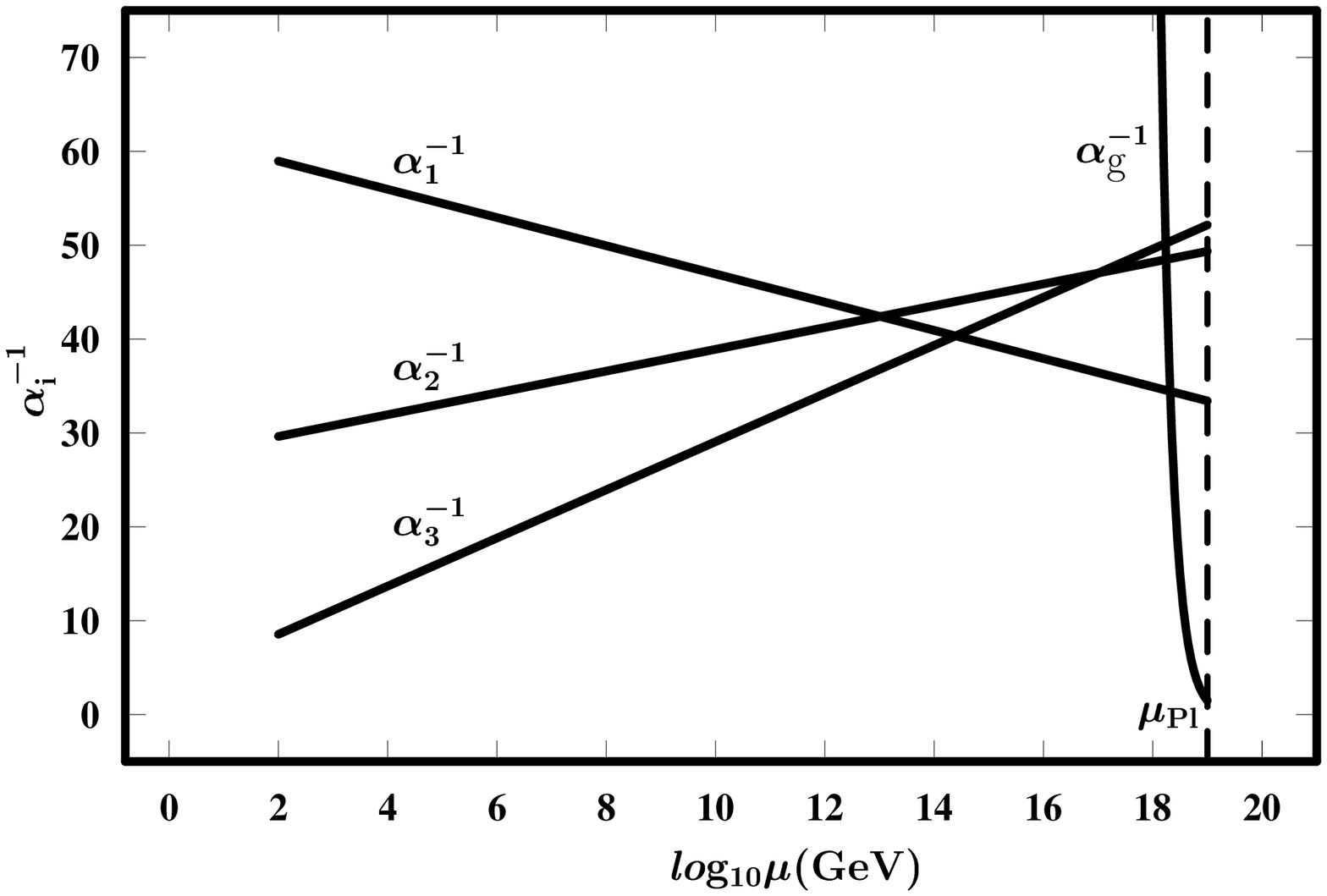}} \caption{}
\end{figure}

In this connection, it is very attractive to include gravity. The
quantity:
\begin{equation}
  \alpha_g = \left(\frac{\mu}{\mu_{Pl}}\right)^2     
\label{2x}
\end{equation}
plays the role of the running "gravitational fine structure
constant" (see Ref.~\cite{Laperashvili:8}) and the evolution 
of its inverse is presented in Fig.~1 together with the evolutions of
$\alpha_i^{-1}(\mu)$.

Assuming the existence of the Dirac relation: $g\tilde g = 2\pi$
for minimal charges, we have the following expression for the
renormalised charges $g$ and $\tilde g$~\cite{Laperashvili:9}:
\begin{equation} 
\alpha(t) \tilde \alpha(t)= \frac{1}{4}.
\end{equation}
Using the Dirac relation, it is easy
to estimate (in the simple SM) the Planck scale value of 
$\tilde\alpha(\mu_{Pl})$ (minimal for $U(1)_Y$ gauge group):
\begin{equation}
\tilde \alpha(\mu_{Pl}) = \frac{5}{3}\alpha_1^{-1}(\mu_{Pl})/4
\approx 55.5/4 \approx 14 \,.                   
\label{3m}
\end{equation} 
This value is really very big compared with the estimate (\ref{2m})
and, of course, with the critical coupling $\tilde\alpha_{crit}\approx1$, 
corresponding to the confinement -- deconfinement phase transition
in the lattice $U(1)$ gauge theory. Clearly we cannot make a
perturbation approximation with such a strong coupling
$\tilde{\alpha}$. It is hard for such monopoles not to be confined.

There is an interesting way out of this problem if one wants to have
the existence of monopoles, namely to extend the SM gauge group so
cleverly that certain selected linear combinations of charges get
bigger electric couplings than the corresponding SM couplings. That
could make the monopoles which, for these certain linear combinations
of charges, couple more weakly and thus have a better chance of being
allowed ``to exist''.

An example of such an extension of the SM that can impose the
possibility of allowing the existence of free monopoles is just
Family Replicated Gauge Group Model (FRGGM).

{}FRGGs of type $[SU(N)]^{N_{fam}}$ lead to the lowering of the
magnetic charge of the monopole belonging to one family:
\begin{equation}
\tilde \alpha_{one\,\,family} = \frac{\tilde \alpha}{N_{fam}}.
\end{equation} 
{}For $N_{fam} = 3$,  for $[SU(2)]^3$ and $[SU(3)]^3$, we have: 
$\tilde \alpha_{one\,\,family}^{(2,3)}={\tilde\alpha}^{(2,3)}/3$. 
{}For the family replicated group
$[U(1)]^{N_{fam}}$ we obtain:
\begin{equation}
\tilde \alpha_{one\,\,family} = \frac{\tilde \alpha}{N^{*}} \, 
\end{equation}
where $N^{*}=\frac{1}{2}N_{fam}(N_{fam}+1)$. For $N_{fam}=3$ 
and $[U(1)]^3$, we have: 
$\tilde \alpha_{\rm one\,\,family}^{(1)}= {\tilde \alpha}^{(1)}/6$ 
(six times smaller!). This result
was obtained previously in Ref.~\cite{Laperashvili:10}.

According to the FRGGM, at some point $\mu=\mu_G<\mu_{Pl}$ (or
really in a couple of steps) the fundamental group 
$G\equiv G_{\mbox{ext}}$ undergoes spontaneous breakdown to its diagonal
subgroup:
\begin{equation}
  G \longrightarrow G_{diag. subgr.} = \{ g, g, g || g \in SMG\},
\label{9m}
\end{equation}
which is identified with the usual (low-energy) group SMG.

In the Anti-GUT-model~\cite{Laperashvili:2,Laperashvili:3} 
the FRGG breakdown was considered at
$\mu_G\sim 10^{18}$ GeV. But the aim of this investigation is to
show that we can see quite different consequences of the extension
of the SM to FRGGM, if the $G$-group undergoes the breakdown to
its diagonal subgroup (that is, SMG) not at $ \mu_G\sim 10^{18}$ 
GeV, but at $\mu_G\sim 10^{14}$ or $ 10^{15}$ GeV, 
\mbox{\it i.e.} before
the intersection of $\alpha_{2}^{-1}(\mu)$ with
$\alpha_{3}^{-1}(\mu)$ at $\mu\approx 10^{16}$ GeV. In this case,
in the region $\mu_G<\mu<\mu_{Pl}$ there are three 
$SMG\times U(1)_{B-L}$ groups for the three FRGG families, and we have a lot
of fermions, mass protected or not mass protected, belonging to
usual families or to mirror ones. In the FRGGM the additional 5
Higgs bosons, with their large VEVs, are responsible for the mass
protection of a lot of new fermions appearing in the region 
$\mu>\mu_G$. Here we denote the total number of fermions $N_F$, which
is different to $N_{fam}$.

Also the role of monopoles can be important in the vicinity of the
Planck scale: they give contributions to the beta-functions and
change the evolution of the $\alpha^{-1}(\mu)$. Finally, we
obtain the following RGEs:
\begin{equation}
   \frac{d(\alpha_i^{-1}(\mu))}{dt} = \frac{b_i}{4\pi } +
   \frac{N_M^{(i)}}{\alpha_i}\beta^{(m)}(\tilde \alpha_{U(1)}) \,  
\label{G3m}
\end{equation}
where $b_i$ are given by the following values:
\begin{eqnarray}
 b_i\!&=&\! (b_1, b_2, b_3)\nonumber \\
 \!&=&\! (-\frac{4N_F}{3} -\frac{1}{10}N_S,\quad
      \frac{22}{3}N_V - \frac{4N_F}{3} -\frac{1}{6}N_S,\quad
      11 N_V - \frac{4N_F}{3})\,.
\label{G4m}
\end{eqnarray}
The integers $N_F,\,N_S,\,N_V,\,N_M\,$ are respectively the
total numbers of fermions, Higgs bosons, vector gauge fields and
scalar monopoles in the FRGGM considered in our theory. In 
our FRGG model we have $N_V=3$, because we have 3 times more gauge
fields $(N_{fam}=3)$, in comparison with the SM and one Higgs
scalar monopole in each family.

We have obtained the evolutions of $\alpha_i^{-1}(\mu)$ near the
Planck scale by numerical calculations for: $\mu_G=10^{14}\,\,$ GeV, 
$N_F=18$, $N_S=6$, $N_M^{(1)}=6$, 
$N_M^{(2,3)}=3$. Fig.~2 shows the existence of the unification point.

\begin{figure}[t]
\centerline{\epsfxsize=\textwidth \epsfbox{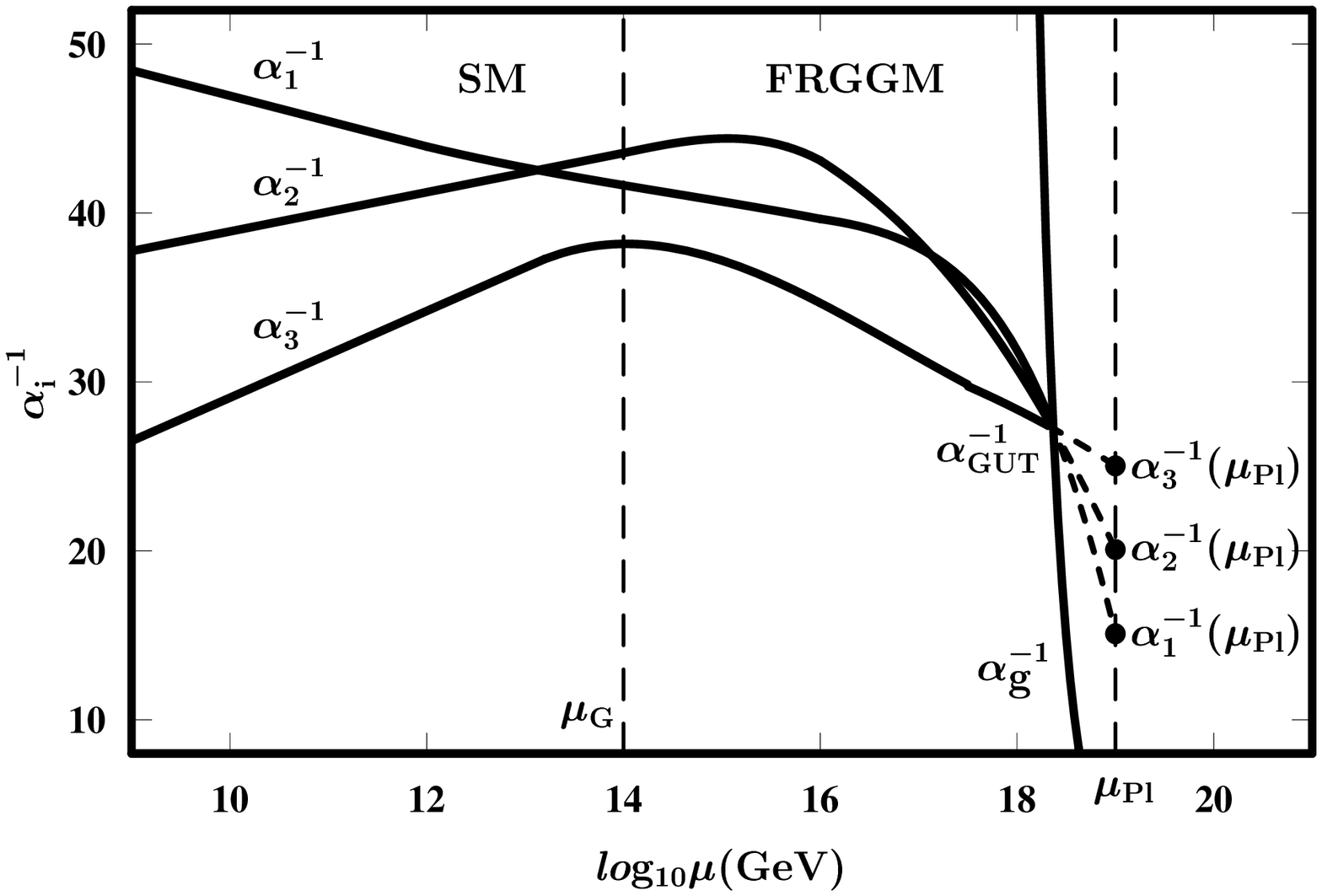}} \caption{}
\end{figure}

We see that in the region $\mu>\mu_G$ a lot of new fermions, and
a number of monopoles near the Planck scale, change the one-loop
approximation behaviour of $\alpha_i^{-1}(\mu)$ which we had in
the SM. In the vicinity of the Planck scale these evolutions begin
to decrease, as the Planck scale $\mu=\mu_{Pl}$ is approached,
implying the suppression of asymptotic freedom in the non-Abelian
theories. Fig.~2 gives the following Planck scale values for the
$\alpha_i$:
\begin{equation}
   \alpha_1^{-1}(\mu_{Pl})\approx 13\,\quad
   \alpha_2^{-1}(\mu_{Pl})\approx 19\,\quad
\alpha_3^{-1}(\mu_{Pl})\approx 24 \,.  \label{G11m} 
\end{equation} 
{}Fig.~2 demonstrates the unification of all gauge interactions, including
gravity (the intersection of $\alpha_g^{-1}$ with $
\alpha_i^{-1}$), at
\begin{equation}
\alpha_{GUT}^{-1}\approx 27 \quad {\rm and} \quad x_{GUT}\approx 18.4 \,.
\end{equation} 
Here we can expect the existence of $[SU(5)]^3$ or 
$[SO(10)]^3$ (SUSY or not SUSY) unification.

Considering the predictions of such a theory for low-energy
physics and cosmology, maybe in future we shall be able to answer
the question: Does the unification $[SU(5)]^3$ or
$[SO(10)]^3$ really exist near the Planck scale?

Recently F.~S.~Ling and P.~Ramond~\cite{Laperashvili:11} 
considered the group of
symmetry $[SO(10)]^3$ and showed that it explains the observed
hierarchies of fermion masses and mixings.

\subsection*{Acknowledgements}

This investigation was supported by the grant RFBR 02-02-17379.


\LastPageEnding
\end{document}